\documentclass[twocolumn,journal]{IEEEtran}
\usepackage[latin9]{inputenc}
\usepackage{microtype}
\usepackage{listings}
\usepackage[htt]{hyphenat}
\usepackage{array}
\usepackage{float}
\usepackage{multirow}
\usepackage{amsmath}
\usepackage{amssymb}
\usepackage{graphicx}
\usepackage{color}
\usepackage{xcolor}
\usepackage{array}
\usepackage{caption}
\usepackage{tabularx}
\usepackage{makecell}
\usepackage{afterpage}
\usepackage{adjustbox}
\usepackage{placeins}
\usepackage{balance}
\usepackage{booktabs}   
\usepackage{siunitx}    
\usepackage{threeparttable} 
\usepackage[labelsep=period]{caption}
\usepackage{tikz} 
\usepackage{stfloats}
\usetikzlibrary{positioning, shapes, fit, calc, arrows, arrows.meta, shapes.geometric}
\usepackage{tcolorbox} 

\newtcolorbox{promptbox}{
  colback=gray!5!white,    
  colframe=gray!15!white,  
  arc=8pt,                 
  boxrule=1pt,             
  left=15pt,               
  right=15pt,              
  top=15pt,                
  bottom=15pt              
}

\usepackage[unicode=true,
 bookmarks=true,bookmarksnumbered=true,bookmarksopen=true,bookmarksopenlevel=1,
 breaklinks=false,pdfborder={0 0 0},pdfborderstyle={},backref=false,colorlinks=false]
 {hyperref}
\hypersetup{pdftitle={Your Title},
 pdfauthor={Your Name},
 pdfpagelayout=OneColumn, pdfnewwindow=true, pdfstartview=XYZ, plainpages=false}

\makeatletter

\floatstyle{ruled}
\newfloat{algorithm}{tbp}{loa}
\providecommand{\algorithmname}{Algorithm}
\floatname{algorithm}{\protect\algorithmname}

\let\oldforeign@language\foreign@language
\DeclareRobustCommand{\foreign@language}[1]{%
  \lowercase{\oldforeign@language{#1}}}

\usepackage[caption=false,font=footnotesize]{subfig}
\usepackage[noadjust]{cite}

\makeatother

\newcommand{\midnotesmall}{\fontsize{8.2pt}{9.8pt}\selectfont}

\renewcommand\arraystretch{1.35} 

\newcolumntype{C}{>{\centering\arraybackslash}X}   
\newcolumntype{M}[1]{>{\centering\arraybackslash}m{#1}} 

\begin{document}
\title{\huge{AutoDP-LLM: Automating Data Pre-processing for Intrusion Detection Systems using Large Language Models}}
\author{Bao-Phong Nguyen, Gia-Khanh Pham, Thai-Duong Do, Mai Xuan Trang, Minh-Tuan Le, \\ Xuan-Nam Tran, Huan Vu, Tien-Cuong Nguyen, Vu-Duc Ngo, and Thien Van Luong

\thanks{Bao-Phong Nguyen, Gia-Khanh Pham, Thai-Duong Do, Huan Vu, and Thien Van Luong are with Business AI Lab, College of Technology, National Economics University, Hanoi, Vietnam (e-mail: \{phongnb.bai, khanhpg.bai, duongdt.bai\}@st.neu.edu.vn, \{huanv, thienlv\}@neu.edu.vn).
} 

\thanks{Mai Xuan Trang is with A2I Lab, School of Computing, Phenikaa University, Hanoi, Vietnam (e-mail: trang.maixuan@phenikaa-uni.edu.vn).
} 

\thanks{Minh-Tuan Le is with the Faculty of Electrical and Electronic 1, Posts and Telecommunications Institute of Technology, Hanoi, Vietnam  (e-mail: tuanlm@ptit.edu.vn).
}

\thanks{Xuan-Nam Tran is with Advanced Wireless Communications Group, Le Quy Don Technical University, Hanoi, Vietnam (e-mail: namtx@lqdtu.edu.vn).
}

\thanks{
Tien-Cuong Nguyen is with VNPT AI, VNPT Group, Hanoi, Vietnam (e-mail: nguyentiencuong@vnpt.vn).}  

\thanks{
Vu-Duc Ngo is with MobiFone HighTech Center, MobiFone Corp., Hanoi, Vietnam (e-mail: duc.ngo@mobifone.vn).}  

\thanks{Corresponding author: Thien Van Luong (e-mail: thienlv@neu.edu.vn)}

}

\markboth{}{}
\maketitle
{
\begin{abstract}
The increasing complexity and scale of modern cyber-attacks demand intelligent and computationally efficient Intrusion Detection Systems (IDS). However, designing effective data pre-processing pipelines traditionally involves substantial trial-and-error effort and repeated evaluation of alternative configurations. For large, high-dimensional network traffic data, this process can create a significant computational burden. In this work, we propose AutoDP-LLM, an automated pre-processing framework designed to reduce manual pipeline development and computational overhead. Specifically, AutoDP-LLM leverages Large Language Models (LLMs) to autonomously generate and validate executable data pre-processing pipelines. The framework combines deterministic host-side planning with LLM-based specialist agents to formulate data-processing strategies, synthesize executable code, and adaptively determine retained feature sets using semantic reasoning and training-derived statistical evidence, without requiring a predefined feature budget. Focusing on multiclass intrusion detection, we evaluate AutoDP-LLM on the UNSW-NB15 and NSL-KDD benchmark datasets using multiple downstream classifiers. Comparative experiments against conventional feature-selection methods show that AutoDP-LLM achieves competitive detection performance while automating the generation of compact and executable pre-processing pipelines. Component-level ablation experiments further demonstrate the complementary contributions of the semantic and statistical feature-reduction components. The repeated generation, validation, execution, and assessment of candidate pipelines are amenable to parallel execution, highlighting the potential of scalable computing environments, including high-performance computing (HPC) systems, to support automated IDS pipeline development.
\end{abstract}
}
    

\section{Introduction}
In the contemporary digital landscape, the increasing complexity and scale of modern cyber-attacks have made cybersecurity an increasingly critical concern, requiring intelligent and computationally efficient Intrusion Detection Systems (IDS). IDS serve as an important defense mechanism by monitoring network traffic and system activities to identify anomalous behavior and potential security threats~\cite{Scarfone2007}. However, as cyber threats become increasingly sophisticated, particularly with the emergence of zero-day exploits and previously unseen attack patterns, traditional signature-based IDS may be insufficient to provide effective detection. Consequently, research has increasingly shifted toward machine learning (ML)-based IDS approaches, which have shown promising capabilities for detecting novel and complex attack patterns~\cite{GarciaTeodoro2009}.

Despite their potential, the effectiveness of ML-based IDS models depends strongly on the quality of input data. Benchmark intrusion detection datasets, such as UNSW-NB15 and NSL-KDD, are often high-dimensional and heterogeneous, containing both numerical and categorical features~\cite{moustafa2015unsw}~\cite{tavallaee2009detailed}. Data pre-processing, including cleaning, normalization, encoding, feature engineering, and feature selection, therefore plays an important role in improving the quality and usability of the data for downstream learning models. However, designing effective data pre-processing pipelines for IDS remains a labor-intensive and expertise-driven task~\cite{Tsai2009}. In practice, identifying suitable data transformations, feature subsets, and parameter configurations often involves repeated trial-and-error exploration. For large-scale and high-dimensional network traffic datasets, the repeated generation, execution, and evaluation of multiple candidate pipelines may introduce considerable computational overhead and scalability challenges, particularly when numerous alternative configurations must be assessed.
Additionally, pre-processing pipelines are often dataset-dependent, and a pipeline developed for one dataset may not generalize well to another dataset with different traffic characteristics, feature distributions, or attack compositions. As a result, applying existing pipelines to new datasets often requires additional manual redesign and tuning effort. Furthermore, suboptimal pre-processing decisions may negatively affect IDS performance, potentially increasing false alarms or reducing attack detection capability. Recent advances in Large Language Models (LLMs) have demonstrated promising capabilities in contextual reasoning, task adaptation, and automated code generation~\cite{Ahmad2024, narayan2022can}, creating new opportunities for intelligent automation in data engineering tasks.

{
Motivated by these challenges, we propose AutoDP-LLM, an LLM-driven automated pre-processing framework for intrusion detection. The proposed framework analyzes dataset metadata and task descriptions to automatically synthesize executable data transformation and feature-reduction pipelines, thereby reducing manual engineering effort and reliance on repeated trial-and-error design. Furthermore, the repeated generation, validation, execution, and evaluation of candidate pipelines can benefit from scalable computing environments, including parallel, distributed, and HPC-enabled infrastructures, particularly for large-scale IDS datasets or multiple candidate pipelines. The main contributions of this work are summarized as follows:

\begin{itemize}

\item We propose AutoDP-LLM, an automated pre-processing framework that combines deterministic host-side planning with LLM-based specialist agents to generate and validate executable data transformation, resampling, and feature-reduction pipelines for intrusion detection datasets.

\item We develop a three-stage architecture integrating analysis and host planning, transformation and resampling, and feature reduction, together with structured contract validation and repair to ensure that the generated functions satisfy predefined structural and runtime requirements before pipeline execution.

\item We develop an adaptive feature-reduction strategy that integrates semantic feature lineage and pruning with training-derived Random Forest and mutual-information evidence to construct compact feature sets without requiring a predefined feature budget. Component-level ablation experiments further evaluate the contributions of the semantic and statistical components.

\item We conduct a comprehensive evaluation on the UNSW-NB15 and NSL-KDD benchmark datasets under multiclass intrusion detection settings. The evaluation covers multiple downstream classifiers, comparative feature-selection baselines, repeated pipeline-generation runs, and different LLM backbones, considering predictive performance, execution time, performance variability, API latency, and computational overhead. Experimental results show that, under the selected representative classifier for each dataset, AutoDP-LLM achieves higher weighted $F_1$-scores than all reimplemented comparative baselines while maintaining competitive execution time.
\end{itemize}

The rest of this paper is organized as follows. Section~II reviews related work on manual data pre-processing and LLM-based approaches in cybersecurity, outlining the motivation for this work. Section~III presents the proposed AutoDP-LLM framework, including its architecture, automated data pre-processing workflow, feature-reduction method, and downstream evaluation procedure. Section~IV presents the experimental results and discussion, evaluating the proposed framework on the UNSW-NB15 and NSL-KDD datasets. Finally, Section~V concludes the paper and outlines future work.
}

\section{Related Work}
This section reviews the existing literature in three areas relevant to this work: (A) conventional data pre-processing and feature-selection techniques for IDS, (B) LLM-based approaches in cybersecurity, and (C) the motivation for automating IDS data-preprocessing workflows.

\subsection{Conventional Data Pre-processing and Feature Selection for IDS}

The performance of an IDS depends heavily on the quality of the input data. Consequently, extensive research has focused on developing effective preprocessing and feature-selection strategies for benchmark intrusion-detection datasets such as UNSW-NB15~\cite{moustafa2015unsw}. To manage the high dimensionality of modern network-traffic data, conventional preprocessing pipelines frequently incorporate feature-selection techniques to improve learning efficiency and reduce computational complexity.

Feature-selection methods aim to identify informative attributes while reducing redundant or irrelevant features. These approaches are commonly categorized as \textit{filter}, \textit{wrapper}, and \textit{embedded} methods. \textit{Filter} methods rely on statistical criteria, such as Information Gain and correlation measures, to rank feature relevance independently of the classifier. For example, Information Gain has been used to identify informative features from the UNSW-NB15 dataset for lightweight IDS construction~\cite{Ghamry2020}. In contrast, \textit{wrapper} methods iteratively evaluate candidate feature subsets according to the predictive performance of a specific learning algorithm~\cite{Janarthanan2017}. Finally, \textit{embedded} methods integrate feature evaluation directly into the model-training process. Tree-based ensemble algorithms such as Random Forest and XGBoost generate feature-importance estimates during training, which can subsequently be used to guide dimensionality reduction and feature selection~\cite{kasongo2020performance}.

Although these approaches can achieve strong predictive performance, they typically require manual design decisions regarding feature-selection criteria, threshold settings, and model-specific configurations. In addition, identifying an effective preprocessing pipeline often involves iterative experimentation and repeated model evaluation, resulting in nontrivial computational costs. Furthermore, preprocessing strategies optimized for one dataset may not generalize well to different network environments or traffic characteristics, highlighting the need for more automated and scalable preprocessing frameworks~\cite{thakkar2022survey}.

\subsection{LLM-based approaches in cybersecurity}
Recent advances in Large Language Models (LLMs) have enabled a wide range of domain-specific applications through their capabilities in reasoning, code generation, and task-specific content generation. For example, CodeT5+ extends the use of LLMs for code understanding and code generation across multiple software engineering tasks~\cite{Wang2023_CodeT5}. In a different application domain, MeshLLM demonstrates that LLMs can progressively understand and generate complex 3D mesh representations, highlighting their potential for handling structured and specialized data generation problems~\cite{MeshLLM_Ref}. Furthermore, LLMs have also been applied to adaptive learning and optimization tasks. For instance, GPT-based systems have been used to iteratively assess model weaknesses and generate customized training samples to improve mathematical problem-solving performance~\cite{MathTutor_Ref}. Collectively, these studies illustrate the versatility of modern LLMs and suggest their potential for supporting automated workflow generation and optimization tasks in complex application domains.

Leveraging capabilities in reasoning, code generation, and task adaptation, LLMs are increasingly being explored for cybersecurity-related tasks. Recent studies suggest that LLMs can be used to analyze, interpret, and summarize various forms of security-related information, including system logs, network traffic, and threat intelligence reports~\cite{BouHarb2024}. These capabilities can assist analysts in identifying abnormal patterns, prioritizing security alerts, and supporting threat investigation and response activities~\cite{Khan2024}. Furthermore, recent studies have explored the application of LLMs to cybersecurity threat detection, including the identification of previously unseen or zero-day threats, highlighting their potential in intelligent security analysis workflows~\cite{Alsuwaiket2025}.

Beyond analysis, LLMs demonstrate substantial potential for automating routine cybersecurity tasks. For instance, they can generate executable scripts for log parsing~\cite{Ahmad2024}, construct regular expressions to support rule-based threat detection~\cite{Ge2024PentestGPT}, and recommend appropriate actions in response to suspicious events. The capacity of LLMs to interpret high-level security requirements and translate them into functional code positions them as formidable assistants for security engineers. Despite this potential, existing research primarily focuses on utilizing LLMs for interpretive or analytical tasks within cybersecurity workflows. Few studies have examined whether LLMs can autonomously construct comprehensive pre-processing pipelines for IDS. This gap motivates further investigation into leveraging LLMs to support advanced, end-to-end data pre-processing workflows for IDS.

\subsection{Motivations of automatic data pre-processing in IDS}

Despite significant progress in manual pre-processing techniques for IDS~\cite{Tsai2009} and the emerging application of LLMs in cybersecurity and data science~\cite{BouHarb2024, Ahmad2024}, a critical gap remains at the intersection of these domains. Existing pre-processing approaches for IDS are effective but rely heavily on human expertise and dataset-specific adjustments, lacking scalability and adaptability across diverse environments~\cite{Ghamry2020}. Meanwhile, current LLMs-based cybersecurity research predominantly focuses on analytical or explanatory tasks~\cite{Khan2024}, offering limited exploration into how these models can automate the technically demanding stages of data preparation. Although broader data science research demonstrates that LLMs can generate code and construct data pre-processing workflows~\cite{Ahmad2024}, their potential to address the unique requirements of IDS, which include reliability, consistency, and domain-specific optimization, remains largely unexplored. This limitation motivates a comprehensive investigation into whether LLMs can transcend supportive roles to autonomously design end-to-end pre-processing pipelines for IDS. By addressing this gap, our work illustrates the potential of LLMs to facilitate more scalable and consistent data engineering workflows.

\begin{figure*}[t]
    \centering
    \includegraphics[width=\linewidth]{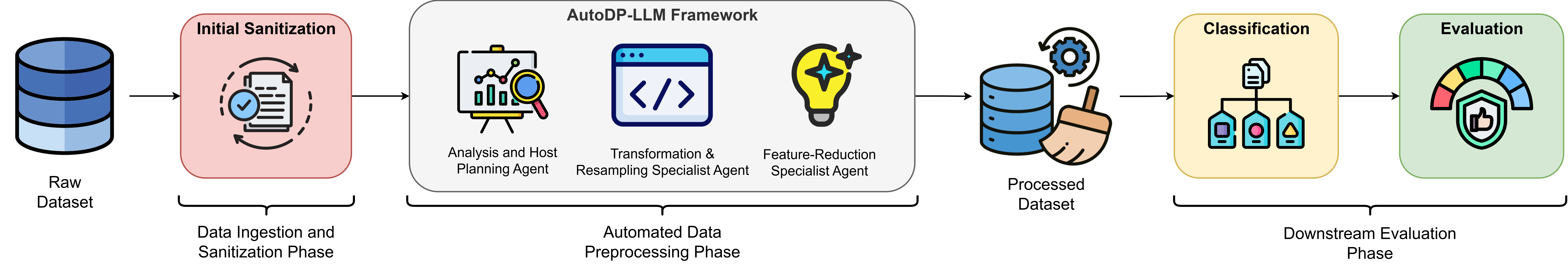}
    \caption{Overall pipeline of the IDS integrating our proposed AutoDP-LLM framework.} 
    \label{fig:figure1}
\end{figure*} 

\section{Proposed AutoDP-LLM Framework}
\label{sec:proposed_framework}

{

This section presents the architecture and operational workflow of the proposed AutoDP-LLM framework. Fig.~\ref{fig:figure1} illustrates the overall three-phase IDS pipeline integrating AutoDP-LLM, whereas Fig.~\ref{fig:figure2} details the three-stage architecture of AutoDP-LLM and its training-data-based preprocessing workflow. The framework comprises three sequential stages: analysis and host planning, transformation and resampling, and feature reduction. These stages integrate deterministic host-side planning with isolated LLM-based specialist agents and structured validation of the generated functions to automate data transformation, class-imbalance handling, and adaptive feature reduction for intrusion detection datasets.

\begin{figure*}[t]
    \centering
    \includegraphics[width=\linewidth]{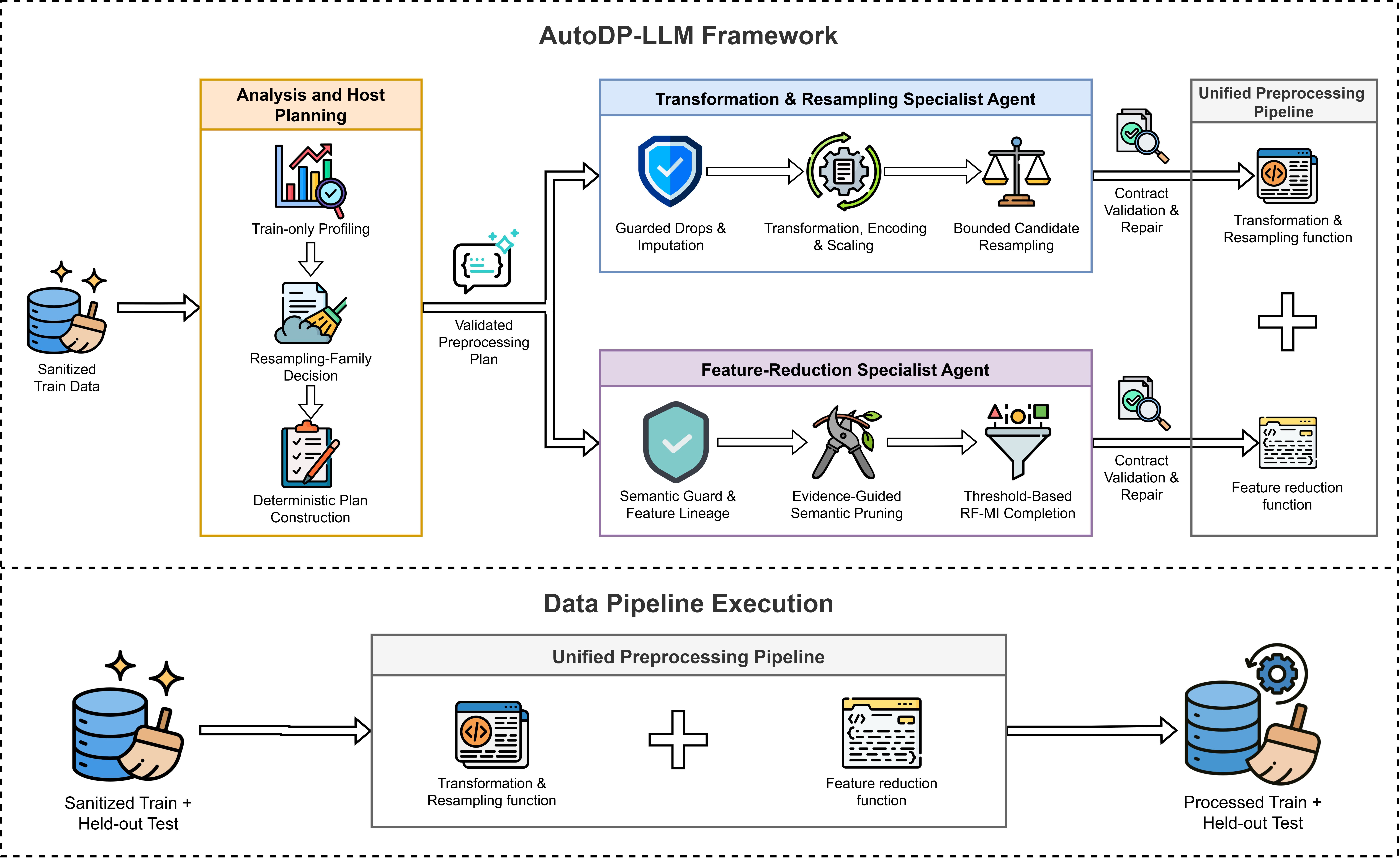}
    \caption{Three-stage architecture and execution workflow of the proposed AutoDP-LLM framework.}
\label{fig:figure2}
\end{figure*}

\subsection{Overall Pipeline Architecture}

As shown in Fig.~\ref{fig:figure1}, the overall IDS pipeline consists of three sequential phases: (1) the \emph{Data Ingestion and Sanitization Phase}, (2) the \emph{Automated Data Preprocessing Phase}, and (3) the \emph{Downstream Evaluation Phase}. The first phase constructs sanitized training and held-out test partitions using dataset-dependent rules. AutoDP-LLM operates within the second phase through three stages: \emph{Analysis and Host Planning}, the \emph{Transformation \& Resampling Specialist Agent}, and the \emph{Feature-Reduction Specialist Agent}. These stages generate and validate the functions that form the \emph{Unified Preprocessing Pipeline}, which produces the \emph{Processed Dataset}. During execution, the adaptive feature set is fixed before the approved resampling candidates are compared using validation data derived from the training partition. The resulting preprocessing configuration is then shared by the downstream classifiers in the third phase.

All learned statistics, transformations, encoders, resamplers, feature-selection evidence, and model-selection decisions are fitted or derived exclusively from the training partition. Held-out test labels, class priors, and feature distributions are unavailable to the planning and selection procedures. The test partition receives only transformations determined from the training partition and is evaluated only after the preprocessing and feature-selection configuration has been fixed.

The modular design allows the same generation and validation workflow to be applied to intrusion detection datasets with different schemas, categorical cardinalities, class distributions, and sample sizes. Dataset-dependent feature descriptions may be provided to support semantic reasoning, but their absence does not prevent statistical feature completion.

\subsection{Data Ingestion and Sanitization Phase}

The \emph{Data Ingestion and Sanitization Phase} comprises two sequential processes: \emph{Data Ingestion} and \emph{Initial Sanitization}. Data ingestion loads the \emph{Raw Dataset} and identifies the dataset-dependent schema and target definition. Initial sanitization then applies deterministic schema normalization and constructs structured training and held-out test partitions. These preliminary operations are implemented using manually defined rules because they address known benchmark conventions rather than data-driven optimization. Typical operations include resolving the target definition and removing fields identified as record identifiers, flow indices, or explicit leakage variables. For multiclass experiments, a redundant binary attack indicator may therefore be removed when the multiclass attack label is designated as the target.

\subsection{Automated Data Preprocessing Phase}

The \emph{Automated Data Preprocessing Phase} comprises the \emph{Analysis and Host Planning}, the \emph{Transformation \& Resampling Specialist Agent}, and the \emph{Feature-Reduction Specialist Agent}. The host derives a train profile from the \emph{Sanitized Train Data} before supplying this profile to the analysis agent. The organization and information flow of these three stages are illustrated in the top panel of Fig.~\ref{fig:figure2}. Each LLM call starts from an isolated conversation with a system prompt designed for its corresponding role and shared constraints. Dataset characteristics, validated handoffs, and implementation requirements are supplied through the corresponding user prompt. The transformation and resampling specialist receives the approved implementation fields and host-defined policy constants, whereas the feature-reduction specialist receives the schema after approved drops and optional feature descriptions. The complete system prompts, execution temperatures, and base contexts of the multi-agent framework are provided in Appendix~\ref{app:system_prompts}.

\subsubsection{Analysis and Host Planning}

The \emph{Analysis and Host Planning} comprises three sequential processes: \emph{Train-only Profiling}, \emph{Resampling-Family Decision}, and \emph{Deterministic Plan Construction}. These processes correspond to the analysis and host planning block in the top panel of Fig.~\ref{fig:figure2}. The complete system prompt governing the analysis agent is provided in Appendix~\ref{subapp:analyst_system}.

First, the host performs this profiling using training data only before the analysis agent is called. Using the sanitized train data, the host summarizes descriptive feature statistics, indicators for missing values and constant features, feature types, categorical cardinalities, class counts and proportions, and dataset size. Relationships between numeric features are summarized using absolute Pearson correlations, with pairs at or above 0.85 included in the prompt. Categorical associations are summarized using bias-corrected Cram\'er's V. For associations between each feature and the target, the host uses one-way ANOVA for numeric features and Cram\'er's V for categorical features, treating the classification target as nominal. These summaries provide contextual evidence for the resampling-family decision but do not define automatic feature-removal rules. Pearson and Spearman correlations are never computed between features and numerically encoded target labels.

Second, the analysis agent receives the train profile and performs the \emph{Resampling-Family Decision}. Its deliberately restricted decision scope permits exactly one choice among no resampling, random oversampling, SMOTE, ADASYN, random undersampling, Tomek links, edited nearest neighbours, SMOTE-Tomek, and SMOTE-ENN. The choice is informed by train class imbalance, absolute class support, categorical cardinalities and the host-defined encoding policy, computational cost, and compatibility with the subsequent preprocessing workflow, which applies the locked Semantic-RF-MI feature set. A large row count alone does not require resampling, and mild imbalance with sufficient class support may lead the agent to retain the original training distribution without resampling. The agent does not generate cleaning decisions, resampling intensities, sampling strategies, row budgets, or a winning candidate. It cannot access validation or test labels, test class priors, or test feature distributions. Consequently, the framework does not attempt to match the label or feature distributions of the training and test partitions.

Third, after the single-field response passes strict JSON validation, the host performs \emph{Deterministic Plan Construction} to produce the \emph{Validated Preprocessing Plan}. The complete schema of the validated preprocessing plan is provided in Appendix~\ref{app:json_schema}. Conservative feature removal is limited to features with all values missing, constant features, exact duplicate features, and identifiers or leakage variables supported by explicitly verified metadata. Feature name keywords alone cannot justify either removal or protection. The host additionally records policies derived from the training data for null handling, categorical encoding, logarithmic transformation, scaling, rare classes, resampling budgets, and candidate selection. When the agent selects no resampling, the host constructs only the \emph{none} candidate. Otherwise, it constructs the \emph{none} reference together with mild and moderate candidates from the selected resampling family. Every candidate must preserve the complete original train class set, including classes representing less than one percent of the data.

For oversampling and hybrid families, a class is eligible only when its support is below one half of the median train class count. For an eligible class~$c$, the unscaled target count is
\begin{equation}
    \widetilde{n}_{c} = \left\lceil n_{c} + \alpha\left(m-n_{c}\right) \right\rceil,
    \label{eq:resampling_target}
\end{equation}
where $n_{c}$ is the original train count of class~$c$, $m$ is the median train class count, and $\alpha$ equals 0.25 and 0.50 for the mild and moderate candidates, respectively. Total expansion is bounded to 1.05 times the original train size for the mild candidate and 1.10 times the original train size for the moderate candidate. If the requested additions exceed the corresponding row budget, they are reduced proportionally and the remaining integer budget is allocated deterministically. Synthetic neighbour methods require at least two original observations in every targeted class. Their neighbour count is set to the smaller of five and the smallest targeted class count minus one. A synthetic family that violates the minimum support requirement is rejected, and the diagnostic recommends random oversampling as an alternative.

For pure undersampling and cleaning families, the mild candidate targets all classes tied for the largest train count, whereas the moderate candidate targets classes with counts strictly above the train median. For a targeted class~$c$, the minimum retained count is
\begin{equation}
    n_c^{\mathrm{floor}} = \left\lceil n_c-\beta(n_c-m) \right\rceil,
    \label{eq:undersampling_floor}
\end{equation}
where $\beta=0.25$ for mild and $\beta=0.50$ for moderate. Random undersampling uses this count as its target, while cleaning methods must retain at least this many observations. Non-target classes retain their original counts. Tomek links permits at most one pass for mild and two passes for moderate. Edited nearest neighbours uses three neighbours for both candidates. The mild candidate removes targeted observations whose labels differ from the modal label of their neighbours, whereas the moderate candidate removes targeted observations if any neighbour has a different label. These policies and the approved preprocessing fields are recorded in the validated preprocessing plan.


\subsubsection{Transformation \& Resampling Specialist Agent}

The \emph{Transformation \& Resampling Specialist Agent} translates the validated preprocessing plan into the \emph{Transformation \& Resampling function}. The structure of the input plan is documented in Appendix~\ref{app:json_schema}, while the complete system prompt governing this specialist agent is provided in Appendix~\ref{subapp:engineer_system}. As shown in Fig.~\ref{fig:figure2}, this agent comprises three sequential processes: (1) \emph{Guarded Drops \& Imputation}, (2) \emph{Transformation, Encoding \& Scaling}, and (3) \emph{Bounded Candidate Resampling}. The function returns the processed train features, their corresponding labels, and the transformed held-out features. All caller-owned inputs are deep-copied before modification, preprocessing quantities are learned from train, and held-out features receive only the corresponding fitted transformations. The selected resampling candidate affects only the final resampling process.

First, \emph{Guarded Drops \& Imputation} applies only feature removals approved in the plan. Numeric infinities are replaced with missing values and imputed using train medians, while categorical missing values are imputed using a deterministic train mode. Rows are not removed to handle missing values, and no imputation statistic is learned from validation or test. Original categorical and numeric source groups are identified before dtype conversion so that categorical values cannot silently pass through fallback numeric coercion.

Second, \emph{Transformation, Encoding \& Scaling} begins by determining eligibility for log transformation from the original train data before resampling. The eligibility snapshot is constructed using only numeric coercion, replacement of infinities with missing values, and imputation with train medians. Eligibility is assessed before clipping, winsorization, outlier treatment, encoding, scaling, or resampling. A numeric feature is eligible for $\log(1+x)$ only when its usable train values are finite and non-negative, it contains at least 20 distinct values, its finite skewness is strictly greater than 2.0, and its upper tail satisfies

\begin{equation}
    \begin{cases}
        Q_{0.99}>Q_{0.75}+3\,\mathrm{IQR}, & \mathrm{IQR}>0,\\
        Q_{0.99}>Q_{0.50}, & \mathrm{IQR}=0,
    \end{cases}
    \label{eq:log_eligibility}
\end{equation}
where $\mathrm{IQR}=Q_{0.75}-Q_{0.25}$ and all quantiles are computed from the train eligibility snapshot. The resulting feature list is applied exactly once to the corresponding train and held-out columns before scaling. Categorical sources and all binary, one-hot, and target-encoded children are excluded from log transformation. Categorical sources are converted to independent string copies and imputed using a deterministic train mode before encoding. Routing is determined from their current train cardinality; the cardinality recorded in the plan serves as reference metadata. A binary source produces one binary column based on a deterministic category order, with unseen held-out categories mapped to the train modal code. Sources with 3--16 categories use one-hot encoding with an explicit unknown category bucket, ensuring that every row activates exactly one child. A conflict between the reserved unknown token and a genuine train category is rejected. Sources with more than 16 categories use fivefold cross-fitting for multiclass target encoding with automatic smoothing, enabled shuffling, and random state 42. The encoder receives the original train label series, and its output width equals the number of original train target classes. Runtime lineage records the encoding route actually used for each source, so one-hot coverage checks apply only to sources processed by one-hot encoding in that invocation. Every categorical source retains explicit lineage from raw features to runtime columns, and all runtime outputs must be numeric. The ordered train and held-out schemas must be identical, and no object, string, categorical, or Boolean types and no missing or infinite values may remain. The function records disjoint runtime column lists for continuous, target-encoded, one-hot, and binary features; together these lists must cover the complete runtime schema. Standard z-score scaling is fitted before resampling using only continuous numeric and target-encoded train columns. Binary and one-hot columns remain unscaled. The same encoding and logarithmic transformation audits are attached to the final train and held-out outputs.

Third, \emph{Bounded Candidate Resampling} executes the approved candidate on train only. The \emph{none} candidate preserves every train row and label. The reference partition is the training partition used to construct the plan, whereas the current partition is the inner training partition processed at runtime. For oversampling and hybrid candidates, approved class targets are transferred from the reference train partition to the current train partition by preserving their lift ratios for each class:
\begin{equation}
    t_c^{\mathrm{run}} = \left\lceil n_c^{\mathrm{run}}\frac{t_c^{\mathrm{ref}}}{n_c^{\mathrm{ref}}} \right\rceil,
    \label{eq:runtime_resampling_target}
\end{equation}
where $n_c^{\mathrm{ref}}$ and $t_c^{\mathrm{ref}}$ are the reference count and approved target, and $n_c^{\mathrm{run}}$ is the current count. The runtime row budget is
\begin{equation}
    B^{\mathrm{run}} = \min\left(\left\lfloor B^{\mathrm{ref}}\frac{N^{\mathrm{run}}}{N^{\mathrm{ref}}} \right\rfloor,\left\lfloor \rho N^{\mathrm{run}} \right\rfloor\right),
    \label{eq:runtime_resampling_budget}
\end{equation}
where $B^{\mathrm{ref}}$ is the approved reference row cap, $N^{\mathrm{ref}}$ and $N^{\mathrm{run}}$ are the reference and current train sizes, and $\rho$ is the approved expansion ratio. Requested additions are reduced proportionally when necessary, with the remaining integer budget allocated deterministically before resampling. For undersampling and cleaning, target classes and retention floors are resolved from the current train counts. Validation and test are never resampled. Output checks cover row-label alignment, schema consistency, finite values, row budgets, and preservation of every original train class; deterministic replay is checked during contract validation.

\subsubsection{Feature-Reduction Specialist Agent}

The \emph{Feature-Reduction Specialist Agent} receives the raw schema after approved drops and optional dataset-dependent feature descriptions and generates the hybrid \emph{Feature reduction function}. The complete system prompt governing this specialist agent is provided in Appendix~\ref{subapp:feature_system}. Following Fig.~\ref{fig:figure2}, this agent comprises three sequential processes: \emph{Semantic Guard \& Feature Lineage}, \emph{Evidence-Guided Semantic Pruning}, and \emph{Threshold-Based RF-MI Completion}. The agent does not receive held-out test statistics, a benchmark-specific mandatory feature list, or the preceding specialist's generated code.

First, \emph{Semantic Guard \& Feature Lineage} establishes at most 20 semantic raw-feature candidates. Each candidate must use an exact name from the supplied schema and be justified by a feature description or an unambiguous IDS meaning. Opaque names are not guessed and remain eligible for statistical completion. Candidates are assigned to one of eight semantic roles covering connection outcomes, protocol or service context, bidirectional traffic magnitude, packet activity, duration or intensity, timing or handshake behaviour, loss or error behaviour, and behavioural aggregation. Each candidate also receives a priority tier: tier~0 for core domain evidence, tier~1 for supporting evidence, and tier~2 for optional evidence. At most 10 tier-0 candidates may be designated as protected. Each nominated raw feature is resolved to its runtime representation. An equal-name runtime column is included when present, and a categorical source is resolved to encoded descendants whose names begin with its exact raw name followed by an underscore. This resolution includes binary, one-hot, and target-encoded descendants. The complete resolved pool is retained for semantic scoring, while every runtime column remains independently eligible for statistical completion.

Second, \emph{Evidence-Guided Semantic Pruning} fits a dedicated random forest on the resolved semantic pool using train data only. This semantic RF uses 50 trees, a maximum depth of eight, class weights inversely proportional to class frequencies and computed separately from each tree's bootstrap sample, and a fixed random seed of 42. The default parallelism setting is retained. For a raw feature group~$g$ with runtime descendants $\mathcal{D}_g$, its group score is
\begin{equation}
    G_g = \sum_{j\in\mathcal{D}_g} I_j^{\mathrm{sem}},
    \label{eq:semantic_group_score}
\end{equation}
where $I_j^{\mathrm{sem}}$ is the importance of runtime column~$j$ in the semantic RF. The representative of each group is its descendant with the highest importance, with ties resolved by column name. A group is discarded when its representative importance is below 0.001. Surviving groups are ranked by ascending tier, protected priority within tier~0, descending group score, and raw feature name. At most the first 15 groups contribute their representatives to the semantic seed. Protected status therefore provides ranking priority but does not exempt a group from the importance threshold. Each selected group contributes exactly one seed column, and the representatives of selected protected groups are recorded as protected anchors.

Third, \emph{Threshold-Based RF-MI Completion} computes independent statistical evidence for every runtime train column. A separate RF fitted using all features uses 100 trees and a maximum depth of ten, with the same class weight, random state, and parallelism settings as the semantic RF. Mutual information uses an explicit type-specific mask for discrete features: binary and one-hot columns are discrete, whereas target-encoded columns are continuous. Other numeric columns are treated as discrete only when their train values justify that interpretation. Every stochastic component uses random state 42. Numerically encoded target identifiers may be supplied to the classifiers and mutual information estimator, but their ordering is never interpreted.
The RF fitted using all features and MI scores are independently converted to percentiles and combined as
\begin{equation}
    S_j = 0.60P_j^{\mathrm{RF}} + 0.40P_j^{\mathrm{MI}},
    \label{eq:rf_mi_score}
\end{equation}
where $P_j^{\mathrm{RF}}$ and $P_j^{\mathrm{MI}}$ denote the percentile scores of runtime column~$j$. Starting from the semantic seed, the selector considers the remaining columns in deterministic score order and admits only candidates with $S_j\geq0.55$. A completion candidate is rejected when its absolute Pearson correlation with any semantic seed column or previously retained completion column is at least 0.90. This filter never removes a semantic seed column, never involves the target, and is not relaxed to reach a desired output size. The fixed thresholds were selected primarily according to their methodological roles. The representative importance threshold of 0.001 removes semantic groups with negligible Random Forest evidence, while the RF-MI threshold of 0.55 requires a moderate level of statistical evidence based on combined percentiles. The correlation threshold of 0.90 removes only strongly redundant candidates while preserving correlated features that may provide complementary IDS information. During method development, limited manual adjustments were made to avoid excessively small or unnecessarily large feature subsets. These adjustments considered only the resulting dimensionality and did not use held-out test labels or downstream test performance. Once established, the thresholds were fixed across all pipeline generation runs and applied unchanged to both datasets. Group level importance aggregation is used only for semantic seed selection. Statistical completion scores individual runtime columns, so additional descendants of a selected group, including a protected group, may enter the final feature set when they independently satisfy the score and redundancy criteria.
The output dimension is determined after completion filtering. If $s$ semantic seed columns and $q$ completion columns are retained, then
\begin{equation}
    K_{\mathrm{final}} = K_{\mathrm{unclamped}} = s+q.
    \label{eq:adaptive_k}
\end{equation}
The selector does not clamp, fill, or truncate this value. The generated function is required to return a nonempty ordered subset of the runtime schema and attach identical feature selection audits to the train and held-out outputs. These audits record the semantic seed, its corresponding raw groups, protected anchors, the complete semantic pool before pruning, the fixed thresholds, and the retained feature counts. The completion count represents selection using combined RF-MI evidence.

\subsubsection{Contract Validation \& Repair}

As illustrated by the \emph{Contract Validation \& Repair} blocks in the top panel of Fig.~\ref{fig:figure2}, this procedure comprises three sequential processes: static validation, isolated runtime validation, and repair feedback.

First, static validation checks the Stage~1 response as a single-field JSON decision and verifies the validated preprocessing plan constructed by the host. The responses from Stages~2 and~3 must each contain one complete Python program and pass import, abstract syntax tree, functional interface, and metadata checks. The policy constants from Stage~2 must match the approved plan, and the generated sampler constructors must belong to the approved family. The Stage~3 checks additionally verify semantic candidate names, role assignments, tier values, the limit of 20 candidates, the limit of 10 protected candidates, and the restriction of protected candidates to tier~0.

Second, isolated runtime validation executes the generated functions on compact fixtures in worker processes with resource limits. The Stage~2 checks cover input immutability, fitting exclusively on training data, categorical lineage, handling of unseen categories, numeric output, scaling, logarithmic transformation eligibility and application, schema consistency, and deterministic replay. Each approved resampling candidate is also exercised using substituted metadata from the compact fixtures. The validator observes the actual sampler calls and checks their strategies, neighbour settings, class-retention constraints, row budgets, and consistency with the unresampled reference.

The Stage~3 runtime checks observe the configurations of the semantic RF and the RF fitted using all features, inspect the discrete-feature mask used by MI, and verify semantic lineage and audit consistency. Dedicated fixtures verify that statistical completion can retain multiple informative descendants of the same raw group. Additional checks examine nominal label invariance, test isolation, deterministic replay, and the rejection of mismatched schemas, inconsistent encoding audits, and nonnumeric columns. The reported feature counts and thresholds are checked against the output schema and the required audit relationships.

Third, a failed response triggers bounded diagnostic feedback and a fresh, isolated LLM call that must regenerate the complete artifact. Stage~1 permits one correction attempt, while Stages~2 and~3 permit two correction attempts each. The initial call uses a temperature of zero, while correction calls use a temperature of 0.1 and increment the generation seed, whose default initial value is 42. Each attempt includes up to three API calls when responses are empty or contain no text. Rejected responses from Stages~2 and~3 and their corresponding diagnostics are retained for inspection. The two accepted programs are combined and checked for syntax validity before the complete module is saved and made available to the unified preprocessing pipeline.

\subsubsection{Data Pipeline Execution}

The bottom panel of Fig.~\ref{fig:figure2} summarizes the \emph{Data Pipeline Execution} of the \emph{Unified Preprocessing Pipeline}. It receives \emph{Sanitized Train + Held-out Test} and produces \emph{Processed Train + Held-out Test}. The execution controller performs three sequential processes: validation splitting from the training partition, locking of the adaptive feature set, and selection among resampling candidates.

First, the sanitized training partition is divided into inner training and validation partitions using a stratified procedure that preserves rare classes, with a validation fraction of 20\% and random state 42. For each class $c$ containing $n_c$ samples, the number assigned to validation is

\begin{equation}
    v_c=\min\left(n_c-1,\left\lfloor0.2n_c\right\rfloor\right).
    \label{eq:rare_safe_validation}
\end{equation}

This rule retains at least one sample from every class in the inner training partition, although extremely rare classes are not necessarily represented in validation. The resulting split is fixed and reused across all pipeline generation runs.

Second, common preprocessing is fitted on the inner training partition and applied to validation. The feature-reduction function is fitted once using the transformed but unresampled \emph{none} candidate, and the resulting adaptive feature set is locked before candidate comparison. For each resampling candidate, resampling is performed on the complete transformed inner training feature matrix before the locked feature set is applied. Validation remains unresampled.

Third, the candidates specified in the validated preprocessing plan are evaluated using a standard LightGBM classifier with random state 42 and verbose output disabled. All remaining parameters retain their default values, and no hyperparameter tuning is performed. The candidate set contains only \emph{none} when the agent selects no resampling; otherwise, it contains \emph{none}, \emph{mild}, and \emph{moderate}. Weighted $F_1$ is used as the primary validation metric. When the weighted $F_1$ difference between two candidates is no greater than 0.003, the candidate with lower intensity is preferred.

The adaptive feature set and resampling winner are selected independently for each successfully generated artifact. Within each artifact, both are subsequently fixed across all downstream classifiers and held-out test processing. The pipeline is not refitted on the complete training partition, and the validation partition is not merged back into the inner training partition.

\subsection{Downstream Evaluation Phase}

As shown in the third phase of Fig.~\ref{fig:figure1}, the \emph{Downstream Evaluation Phase} comprises two sequential processes: \emph{Classification} and \emph{Evaluation}. First, Random Forest, KNN, MLP, XGBoost, LightGBM, and CatBoost are trained on the processed inner training partition using the same locked feature set and resampling winner selected for the corresponding artifact. The LightGBM validation controller is used solely to select among resampling candidates and does not select a separate preprocessing configuration for each downstream classifier.

Second, each trained downstream classifier is evaluated exactly once on the held-out test partition for each successful pipeline generation run. The evaluation reports weighted $F_1$ as the primary predictive metric, together with accuracy, weighted precision, and weighted recall as complementary measures. Operational measurements are reported separately and are not used for pipeline selection or optimization based on held-out test results.
}

\subsection{Scalability Considerations}

Although AutoDP-LLM is evaluated on a single computing platform in this study, several stages of the workflow exhibit natural task-level parallelism. In particular, independently generated preprocessing pipelines can be executed and evaluated separately, while multiple downstream machine-learning models can be trained and assessed concurrently using the resulting feature representations. These independent workloads create opportunities for parallel execution across available computing resources.

Consequently, the proposed workflow is amenable to scalable execution environments, including multicore, distributed, and HPC-enabled computing infrastructures. Such environments may help reduce the overall time required for repeated pipeline generation and evaluation, large-scale experimental studies, or assessments involving multiple datasets and downstream classifiers. While the present study focuses on a single-platform implementation, the identified task-level parallelism motivates future investigations into multicore, distributed, and HPC-enabled implementations of the proposed framework.

{

\section{Experimental Results and Discussion}

\subsection{Overview of Benchmark Datasets}

To evaluate AutoDP-LLM, we conduct multiclass intrusion-detection experiments on two benchmark datasets: UNSW-NB15 and NSL-KDD. For both datasets, we use the predefined training and test splits. All preprocessing steps, including encoding, scaling, and resampling, are fitted only on the training data to avoid data leakage.

\subsubsection{UNSW-NB15 Dataset}
UNSW-NB15~\cite{moustafa2015unsw} contains approximately 2.54 million network-flow records covering normal traffic and nine attack categories: Analysis, Backdoor, DoS, Exploits, Fuzzers, Generic, Reconnaissance, Shellcode, and Worms. We use the predefined split of 175,341 training samples and 82,332 test samples. Each record contains 42 predictor features, a multiclass label (attack\_cat), a binary label (label), and an identifier (id). For multiclass classification, id and label are removed. All ten classes are retained. The dataset is highly imbalanced, particularly for minority classes such as Worms and Backdoor. Placeholder values (``-'') in the service feature are mapped to other.

\subsubsection{NSL-KDD Dataset}
NSL-KDD~\cite{tavallaee2009detailed} is an improved version of KDD Cup 1999 with fewer redundant and duplicate records. We use the predefined training and test sets containing 125,973 and 22,544 samples, respectively. The difficulty score is removed because it is not a network feature, while num\_outbound\_cmds is removed because it has zero variance. This leaves 40 predictor features. Attack labels are grouped into five classes: Normal, DoS, Probe, R2L, and U2R. The categorical features protocol\_type, service, and flag are encoded using mappings learned from the training set, with unseen test categories assigned to an out-of-vocabulary token.
}

\subsection{Implementation Setting}

\subsubsection{Evaluation Metrics}

The performance of the evaluated classifiers is assessed using Precision, Recall, and $F_1\text{-score}$. The $F_1\text{-score}$ is computed as the harmonic mean of Precision and Recall:

\begin{equation}
F_1\text{-score}
=
2 \times
\frac{\text{Precision} \times \text{Recall}}
{\text{Precision} + \text{Recall}}.
\end{equation}

For multiclass classification, weighted averaging is adopted, where the metric for each class is weighted according to its support (the number of true instances belonging to that class).

\subsubsection{Classification Models}

To evaluate the effectiveness of the generated preprocessing pipelines, six machine-learning classifiers are considered. Unless otherwise specified, the default hyperparameters provided by the corresponding software libraries are used:

\begin{itemize}
\item Random Forest (default configuration)
\item K-Nearest Neighbors (KNN) (n\_neighbors = 5)
\item Multi-Layer Perceptron (MLP) (default configuration)
\item XGBoost (default configuration)
\item LightGBM (default configuration)
\item CatBoost (default configuration)
\end{itemize}

\subsubsection{Baseline Methods}

To assess the effectiveness of AutoDP-LLM, we compare it against four literature-based preprocessing or
feature-selection methods:

\begin{itemize}

\item LiteShield-MI-FS: a mutual-information-based feature-selection method \cite{mabulage2026liteshield}.

\item XGBoost-FS: a feature-selection method based on feature-importance scores produced by XGBoost \cite{pansari2024attack}.

\item IGRF-RFE: a hybrid feature-selection method that combines Information-Gain Random Forest and Recursive Feature Elimination (RFE) \cite{yin2023igrf}.

\item DSSTE: a domain-specific statistical transformation and encoding baseline designed for tabular intrusion-detection features \cite{liu2020intrusion}

\end{itemize}

\subsubsection{Experimental Environment}

Experiments were conducted on a workstation equipped with the following hardware and software:

\begin{itemize}
\item CPU: Intel Core i5-13500 (13th Generation)
\item RAM: 64 GB
\item GPU: NVIDIA GeForce RTX 3060 (12 GB VRAM)
\item Operating System: Ubuntu 24.04.4 LTS
\item Software: Python 3.12.3
\item Libraries: scikit-learn (v1.8.0), xgboost (v3.2.0), lightgbm (v4.5.0), and catboost (v1.2.10).
\end{itemize}



\begin{table*}[b]
\centering
\small
\caption{{Classification performance and total execution time on UNSW-NB15 across 30 pipeline-generation runs}}
\label{tab:classifier_comparison_unsw}
\begin{tabularx}{0.95\textwidth}{lCCCCC}
    \toprule
    \textbf{Classifier}
    & \textbf{Accuracy}
    & \textbf{Precision}
    & \textbf{Recall}
    & \textbf{$F_1$-score}
    & \textbf{Execution Time} \\
    \midrule
    Random Forest
    & $76.59 \pm 0.28$
    & $82.64 \pm 0.33$
    & $76.59 \pm 0.28$
    & $78.41 \pm 0.21$
    & $117.18 \pm 9.80$ \\

    KNN
    & $73.33 \pm 0.17$
    & $80.79 \pm 0.33$
    & $73.33 \pm 0.17$
    & $76.08 \pm 0.18$
    & $108.98 \pm 9.77$ \\

    MLP
    & \textbf{77.51 $\pm$ 0.64}
    & $81.17 \pm 0.84$
    & \textbf{77.51 $\pm$ 0.64}
    & $77.76 \pm 0.49$
    & $190.61 \pm 9.30$ \\

    XGBoost
    & $77.20 \pm 0.23$
    & \textbf{82.84 $\pm$ 0.83}
    & $77.20 \pm 0.23$
    & \textbf{78.41 $\pm$ 0.22}
    & \textbf{106.38 $\pm$ 9.77} \\

    LightGBM
    & $77.35 \pm 0.30$
    & $82.82 \pm 0.79$
    & $77.35 \pm 0.30$
    & $78.40 \pm 0.31$
    & $117.91 \pm 9.72$ \\

    CatBoost
    & $76.84 \pm 0.12$
    & $82.46 \pm 0.76$
    & $76.84 \pm 0.12$
    & $77.92 \pm 0.14$
    & $150.56 \pm 9.55$ \\
    \bottomrule
\end{tabularx}
\end{table*}

\vspace{2em}

\begin{table*}[t]
\centering
\small
\caption{Classification performance and total execution time on NSL-KDD across 30 pipeline-generation runs}
\label{tab:classifier_comparison_nsl}
\begin{tabularx}{0.95\textwidth}{lCCCCC}
    \toprule
    \textbf{Classifier}
    & \textbf{Accuracy}
    & \textbf{Precision}
    & \textbf{Recall}
    & \textbf{$F_1$-score}
    & \textbf{Execution Time} \\
    \midrule
    Random Forest
    & $76.46 \pm 0.85$
    & $81.97 \pm 1.00$
    & $76.46 \pm 0.85$
    & $72.43 \pm 0.93$
    & $93.94 \pm 14.59$ \\

    KNN
    & $77.22 \pm 0.52$
    & $80.77 \pm 0.72$
    & $77.22 \pm 0.52$
    & $74.75 \pm 0.50$
    & \textbf{90.97 $\pm$ 14.65} \\

    MLP
    & $78.80 \pm 0.92$
    & $82.63 \pm 0.83$
    & $78.80 \pm 0.92$
    & \textbf{76.38 $\pm$ 0.99}
    & $112.65 \pm 16.28$ \\

    XGBoost
    & $77.79 \pm 0.80$
    & $82.68 \pm 0.63$
    & $77.79 \pm 0.80$
    & $74.37 \pm 0.72$
    & $91.60 \pm 14.57$ \\

    LightGBM
    & $75.34 \pm 0.78$
    & $79.94 \pm 0.33$
    & $75.34 \pm 0.78$
    & $72.07 \pm 0.66$
    & $96.04 \pm 14.61$ \\

    CatBoost
    & \textbf{79.24 $\pm$ 0.26}
    & \textbf{83.82 $\pm$ 0.19}
    & \textbf{79.24 $\pm$ 0.26}
    & $76.13 \pm 0.33$
    & $99.94 \pm 14.61$ \\
    \bottomrule
\end{tabularx}
\end{table*}

\vspace{2em}

\begin{table*}[b]
\centering
\small
\caption{Performance and execution time comparison against other methods under the selected downstream classifiers: XGBoost for UNSW-NB15 and MLP for NSL-KDD}
\label{tab:sota_single_model_comparison_unsw_nsl}
\begin{tabularx}{0.9\textwidth}{lCCCC}
    \toprule
    \multirow{2}{*}{\textbf{Framework / Method}}
    & \multicolumn{2}{c}{\textbf{UNSW-NB15 (XGBoost)}}
    & \multicolumn{2}{c}{\textbf{NSL-KDD (MLP)}} \\
    \cmidrule(lr){2-3}
    \cmidrule(lr){4-5}
    & \textbf{Weighted $F_1$ (\%)}
    & \textbf{Time (s)}
    & \textbf{Weighted $F_1$ (\%)}
    & \textbf{Time (s)} \\
    \midrule

    LiteShield-MI-FS$^\dagger$~\cite{mabulage2026liteshield}
    & 78.17 & 114.07
    & 67.34 & 96.47 \\

    XGBoost-FS$^\dagger$~\cite{pansari2024attack}
    & 77.92 & \textbf{10.74}
    & 71.53 & 78.28 \\

    IGRF-RFE$^\dagger$~\cite{yin2023igrf}
    & 78.13 & 857.20
    & -- & -- \\

    DSSTE$^\dagger$ ~\cite{liu2020intrusion}
    & -- & --
    & 73.59 & \textbf{77.56} \\
    \midrule

    \textbf{AutoDP-LLM (Ours)}
    & \textbf{78.41 $\pm$ 0.22} & 106.38 $\pm$ 9.77
    & \textbf{76.38 $\pm$ 0.99} & $112.65 \pm 16.28$ \\
    \bottomrule

    \multicolumn{5}{l}{
        \footnotesize $^\dagger$ \textit{Re-implemented and reproduced by the authors under the same classifier setting for each dataset.}
    }
\end{tabularx}
\end{table*}

\subsection{Results and Discussion}

{For each experimental configuration, we perform 30 repeated pipeline-generation runs. Each run invokes the LLM anew and produces a separately generated preprocessing-code artifact, which is subsequently validated, executed, and evaluated. To maintain a consistent evaluation protocol across runs, the training-derived validation split, resampling procedures, and stochastic downstream operations use a fixed experimental random state of 42. The LLM invocation settings are detailed in Appendix~\ref{app:system_prompts}, while the contract-repair procedure is described in Section~III. In Tables~\ref{tab:classifier_comparison_unsw} and~\ref{tab:classifier_comparison_nsl}, results are reported as the arithmetic mean $\pm$ standard deviation (SD) across the 30 generated and executed pipeline artifacts. In Table~\ref{tab:sota_single_model_comparison_unsw_nsl}, AutoDP-LLM results use the same mean $\pm$ SD format, whereas the baseline results are single-point estimates under their respective evaluation protocols. The same mean $\pm$ SD convention is used for the component-level ablation study in Table~\ref{tab:feature_reduction_ablation}, the cross-LLM downstream evaluations in Tables~\ref{tab:full_metrics_comparison_unsw} and~\ref{tab:full_metrics_comparison_nsl}, and the operational measures in Table~\ref{tab:llm_backbone_comparison}. The retained-feature entries in Table~\ref{tab:llm_backbone_comparison} are instead reported as the observed minimum--maximum range across runs. The SD summarizes the observed run-to-run dispersion among the separately generated and executed pipelines under the fixed data split and downstream random-state configuration. For percentage-based classification metrics, the SD is expressed in percentage points. For execution time and the operational measures in Table~\ref{tab:llm_backbone_comparison}, the mean and SD are expressed in the units stated in the corresponding column headings.}

{
To ensure clarity in our evaluation tables, we apply specific highlighting rules. In the main performance tables (Tables~\ref{tab:classifier_comparison_unsw} and~\ref{tab:classifier_comparison_nsl}), the values highlighted in \textbf{bold} represent the best performance across all evaluated models. In the baseline comparison (Table~\ref{tab:sota_single_model_comparison_unsw_nsl}), the best-performing method is highlighted in \textbf{bold}. In the component-level ablation study (Table~\ref{tab:feature_reduction_ablation}), the values highlighted in \textbf{bold} represent the highest weighted $F_1$-score for each dataset. In the cross-LLM downstream evaluations (Tables~\ref{tab:full_metrics_comparison_unsw} and~\ref{tab:full_metrics_comparison_nsl}), the values highlighted in \textbf{bold} represent the best performance achieved by each LLM engine, whereas the values highlighted in \textcolor{red}{\textbf{bold and red}} represent the best performance across all three evaluated LLM engines and downstream classifiers. In the operational efficiency comparison (Table~\ref{tab:llm_backbone_comparison}), the values highlighted in \textcolor{red}{\textbf{bold and red}} represent the most favorable result across the three evaluated LLM engines for each dataset, where lower output-token counts, API latency, monetary cost, and training time are considered better. The number of retained features is reported for comparison and is not highlighted as a best-performing result.}

\subsubsection{Performance on Benchmark Datasets}

{Tables~I and~II report the classification performance and total execution time of the AutoDP-LLM pipelines on UNSW-NB15 and NSL-KDD, respectively.}

{On UNSW-NB15, Random Forest and XGBoost both attain the highest mean weighted $F_1$-score at the displayed precision, with $78.41 \pm 0.21\%$ and $78.41 \pm 0.22\%$, respectively. LightGBM follows closely with $78.40 \pm 0.31\%$. MLP produces the highest mean accuracy ($77.51 \pm 0.64\%$), but its weighted $F_1$-score is lower ($77.76 \pm 0.49\%$), indicating that the classifier ranking depends on the evaluation metric. XGBoost records the lowest mean total execution time ($106.38 \pm 9.77$~s), whereas MLP records the highest ($190.61 \pm 9.30$~s). XGBoost combines one of the highest weighted $F_1$-scores with the lowest mean execution time on UNSW-NB15.}

{On NSL-KDD, MLP achieves the highest mean weighted $F_1$-score, $76.38 \pm 0.99\%$, followed by CatBoost at $76.13 \pm 0.33\%$. CatBoost attains the highest mean accuracy ($79.24 \pm 0.26\%$) and has a lower mean execution time than MLP ($99.94 \pm 14.61$~s versus $112.65 \pm 16.28$~s). Their mean weighted $F_1$-scores differ by only 0.25 percentage points, while CatBoost exhibits lower observed dispersion. KNN has the lowest mean execution time ($90.97 \pm 14.65$~s), although its weighted $F_1$-score is 1.63 percentage points below that of MLP.}

{Across the two datasets, the selected and unclamped feature counts remain within 26--28 features for UNSW-NB15 and 24--26 features for NSL-KDD. The framework therefore produces compact representations under two different schemas and class distributions.}

\begin{table*}[t]
\centering
\midnotesmall
\caption{Component configurations and weighted $F_1$-scores of the feature-reduction ablation study}
\label{tab:feature_reduction_ablation}
\begin{tabularx}{0.98\textwidth}{lCCCCM{2.8cm}M{2.8cm}}
    \toprule
    \multirow{2}{*}{\textbf{Ablation Variant}}
    & \multicolumn{3}{c}{\textbf{Component Configuration}}
    & \multicolumn{2}{c}{\textbf{Results (Weighted $F_1$})} \\
    \cmidrule(lr){2-4}
    \cmidrule(lr){5-6}
    & \shortstack{\textbf{Semantic Guard}}
    & \shortstack{\textbf{Semantic Pruning}}
    & \shortstack{\textbf{RF-MI Completion}}
    & \shortstack{\textbf{UNSW-NB15}}
    & \shortstack{\textbf{NSL-KDD}} \\
    \midrule
    w/o RF-MI Completion
    & $\checkmark$
    & $\checkmark$
    & 
    & 77.28 $\pm$ 0.17
    & 75.03 $\pm$ 0.47 \\

    w/o Semantic Pruning
    & $\checkmark$
    & 
    & $\checkmark$
    & 77.90 $\pm$ 0.10
    & 76.26 $\pm$ 1.14 \\

    w/o Semantic Components
    & 
    & 
    & $\checkmark$
    & 78.33 $\pm$ 0.00
    & 74.53 $\pm$ 0.00 \\
    
    Full method
    & $\checkmark$
    & $\checkmark$
    & $\checkmark$
    & \textbf{78.41 $\pm$ 0.22}
    & \textbf{76.38 $\pm$ 0.99} \\

    \bottomrule
    \multicolumn{6}{l}{\footnotesize XGBoost and MLP are used as the fixed representative classifiers for UNSW-NB15 and NSL-KDD, respectively.}
\end{tabularx}
\end{table*}


\begin{table*}[b]
  \centering
  \small
  \caption{Classification performance of the evaluated LLM engines on UNSW-NB15.}
  \label{tab:full_metrics_comparison_unsw}
  \begin{tabularx}{0.95\textwidth}{llCCCC}
    \toprule
    \multirow{2}{*}{\textbf{LLM Engine}}
    & \multirow{2}{*}{\textbf{Classifier}}
    & \multicolumn{4}{c}{\textbf{Classification Performance}} \\
    \cmidrule(lr){3-6}
    & & \textbf{Accuracy}
      & \textbf{Precision}
      & \textbf{Recall}
      & \textbf{$F_1$-score} \\
    \midrule

    \multirow{3}{*}{Claude-Sonnet-4.6}
      & Random Forest & 74.96 $\pm$ 0.31
                      & \textcolor{red}{\textbf{83.38 $\pm$ 0.38}}
                      & 74.96 $\pm$ 0.31
                      & \textbf{77.74 $\pm$ 0.15} \\
      & XGBoost       & 74.84 $\pm$ 0.25
                      & 83.20 $\pm$ 0.48
                      & 74.84 $\pm$ 0.25
                      & 77.44 $\pm$ 0.18 \\
      & MLP           & \textbf{76.19 $\pm$ 0.95}
                      & 81.47 $\pm$ 0.48
                      & \textbf{76.19 $\pm$ 0.95}
                      & 77.13 $\pm$ 0.48 \\
    \midrule

    \multirow{3}{*}{DeepSeek-V4-Pro}
      & Random Forest & 76.84 $\pm$ 0.17
                      & \textbf{82.31 $\pm$ 0.08}
                      & 76.84 $\pm$ 0.17
                      & \textcolor{red}{\textbf{78.56 $\pm$ 0.08}} \\
      & XGBoost       & \textbf{77.23 $\pm$ 0.18}
                      & 81.83 $\pm$ 0.13
                      & \textbf{77.23 $\pm$ 0.18}
                      & 78.34 $\pm$ 0.11 \\
      & MLP           & 75.96 $\pm$ 1.22
                      & 80.60 $\pm$ 0.59
                      & 75.96 $\pm$ 1.22
                      & 76.78 $\pm$ 0.70 \\
    \midrule

    \multirow{3}{*}{GPT-5.4}
      & Random Forest & 76.59 $\pm$ 0.28
                      & 82.64 $\pm$ 0.33
                      & 76.59 $\pm$ 0.28
                      & \textbf{78.41 $\pm$ 0.21} \\
      & XGBoost       & 77.20 $\pm$ 0.23
                      & \textbf{82.84 $\pm$ 0.83}
                      & 77.20 $\pm$ 0.23
                      & \textbf{78.41 $\pm$ 0.22} \\
      & MLP           & \textcolor{red}{\textbf{77.51 $\pm$ 0.64}}
                      & 81.17 $\pm$ 0.84
                      & \textcolor{red}{\textbf{77.51 $\pm$ 0.64}}
                      & 77.76 $\pm$ 0.49 \\
    \bottomrule
  \end{tabularx}
\end{table*}


\begin{table*}[b]
  \centering
  \small
  \caption{Classification performance of the evaluated LLM engines on NSL-KDD.}
  \label{tab:full_metrics_comparison_nsl}
  \begin{tabularx}{0.95\textwidth}{llCCCC}
    \toprule
    \multirow{2}{*}{\textbf{LLM Engine}}
    & \multirow{2}{*}{\textbf{Classifier}}
    & \multicolumn{4}{c}{\textbf{Classification Performance}} \\
    \cmidrule(lr){3-6}
    & & \textbf{Accuracy}
      & \textbf{Precision}
      & \textbf{Recall}
      & \textbf{$F_1$-score} \\
    \midrule

    \multirow{3}{*}{Claude-Sonnet-4.6}
      & Random Forest & 76.42 $\pm$ 0.05
                      & 82.33 $\pm$ 0.03
                      & 76.42 $\pm$ 0.05
                      & 72.81 $\pm$ 0.10 \\
      & XGBoost       & \textbf{78.67 $\pm$ 0.42}
                      & \textcolor{red}{\textbf{83.30 $\pm$ 0.22}}
                      & \textbf{78.67 $\pm$ 0.42}
                      & \textbf{76.05 $\pm$ 0.41} \\
      & MLP           & 76.87 $\pm$ 0.26
                      & 79.73 $\pm$ 1.01
                      & 76.87 $\pm$ 0.26
                      & 74.82 $\pm$ 0.40 \\
    \midrule

    \multirow{3}{*}{DeepSeek-V4-Pro}
      & Random Forest & 76.35 $\pm$ 0.63
                      & 82.13 $\pm$ 0.44
                      & 76.35 $\pm$ 0.63
                      & 72.41 $\pm$ 0.55 \\
      & XGBoost       & \textbf{77.31 $\pm$ 0.85}
                      & \textbf{82.43 $\pm$ 0.33}
                      & \textbf{77.31 $\pm$ 0.85}
                      & 73.94 $\pm$ 1.00 \\
      & MLP           & 77.28 $\pm$ 1.15
                      & 81.45 $\pm$ 0.96
                      & 77.28 $\pm$ 1.15
                      & \textbf{75.19 $\pm$ 1.10} \\
    \midrule

    \multirow{3}{*}{GPT-5.4}
      & Random Forest & 76.46 $\pm$ 0.85
                      & 81.97 $\pm$ 1.00
                      & 76.46 $\pm$ 0.85
                      & 72.43 $\pm$ 0.93 \\
      & XGBoost       & 77.79 $\pm$ 0.80
                      & \textbf{82.68 $\pm$ 0.63}
                      & 77.79 $\pm$ 0.80
                      & 74.37 $\pm$ 0.72 \\
      & MLP           & \textcolor{red}{\textbf{78.80 $\pm$ 0.92}}
                      & 82.63 $\pm$ 0.83
                      & \textcolor{red}{\textbf{78.80 $\pm$ 0.92}}
                      & \textcolor{red}{\textbf{76.38 $\pm$ 0.99}} \\
    \bottomrule
  \end{tabularx}
\end{table*}

\subsubsection{Comparison with Baseline Methods}

{Table~\ref{tab:sota_single_model_comparison_unsw_nsl} compares AutoDP-LLM with the re-implemented feature-selection baselines under the respective fixed downstream classifiers: XGBoost for UNSW-NB15 and MLP for NSL-KDD. Consequently, the differences below are interpreted descriptively.}

{On UNSW-NB15, AutoDP-LLM obtains a mean weighted $F_1$-score of $78.41 \pm 0.22\%$. The corresponding point estimates are $78.17\%$ for LiteShield-MI-FS, $78.13\%$ for IGRF-RFE, and $77.92\%$ for XGBoost-FS. AutoDP-LLM therefore has the highest reported weighted-$F_1$ value, with its mean exceeding the strongest listed baseline point estimate by 0.24 percentage points. Its mean execution time is $106.38 \pm 9.77$~s, compared with $114.07$~s for LiteShield-MI-FS, $10.74$~s for XGBoost-FS, and $857.20$~s for IGRF-RFE. Accordingly, AutoDP-LLM is faster than LiteShield-MI-FS and IGRF-RFE in this comparison, but it does not match the substantially lower execution time of XGBoost-FS.}

{On NSL-KDD, AutoDP-LLM achieves a mean weighted $F_1$-score of $76.38 \pm 0.99\%$, compared with $73.59\%$ for DSSTE, $71.53\%$ for XGBoost-FS, and $67.34\%$ for LiteShield-MI-FS. Its mean is 2.79 percentage points above the strongest listed baseline point estimate. However, its mean execution time of $112.65 \pm 16.28$~s is higher than those of LiteShield-MI-FS ($96.47$~s), XGBoost-FS ($78.28$~s), and DSSTE ($77.56$~s). The comparison therefore indicates a predictive-performance advantage at the point-estimate level on NSL-KDD, accompanied by greater execution cost.}

{Overall, AutoDP-LLM reports the highest weighted-$F_1$ value in the comparison on both datasets, but it does not provide a consistent runtime advantage.}

\begin{table*}[t]
    \centering
    \small
    \caption{Operational efficiency of the evaluated LLM engines across datasets}
    \label{tab:llm_backbone_comparison}
    \begin{tabularx}{0.95\textwidth}{lCCCCC}
        \toprule
        & \multicolumn{3}{c}{\textbf{Design-time Overhead}}
        & \multicolumn{2}{c}{\textbf{Run-time Efficiency}} \\
        \cmidrule(lr){2-4}
        \cmidrule(lr){5-6}
        \textbf{LLM Engine}
        & \textbf{Output Tokens}
        & \textbf{API Latency (s)}
        & \textbf{Cost (USD)}
        & \textbf{Number of Features}
        & \textbf{Training Time (s)} \\
        \midrule

        \multicolumn{6}{c}{
            \textbf{Dataset: UNSW-NB15
            \,
            \textit{(Representative classifier: XGBoost)}}
        } \\
        \midrule

        Claude-Sonnet-4.6
        & 18,199 $\pm$ 11,226
        & 283.26 $\pm$ 99.65
        & \$0.318 $\pm$ 0.168
        & $26\,\text{--}\,27$
        & 14.73 $\pm$ 0.46 \\

        GPT-5.4
        & \textcolor{red}{\textbf{10,418 $\pm$ 2,139}}
        & \textcolor{red}{\textbf{60.23 $\pm$ 8.44}}
        & \$0.194 $\pm$ 0.032
        & $26\,\text{--}\,28$
        & \textcolor{red}{\textbf{13.55 $\pm$ 0.45}} \\

        DeepSeek-V4-Pro
        & 79,152 $\pm$ 33,595
        & 858.33 $\pm$ 354.50
        & \textcolor{red}{\textbf{\$0.147 $\pm$ 0.059}}
        & $27\,\text{--}\,29$
        & 14.19 $\pm$ 0.42 \\
        \midrule

        \multicolumn{6}{c}{
            \textbf{Dataset: NSL-KDD
            \,
            \textit{(Representative classifier: MLP)}}
        } \\
        \midrule

        Claude-Sonnet-4.6
        & 13,585 $\pm$ 8,422
        & 306.32 $\pm$ 74.46
        & \$0.249 $\pm$ 0.126
        & $24\,\text{--}\,27$
        & 12.67 $\pm$ 0.40 \\

        GPT-5.4
        & \textcolor{red}{\textbf{10,800 $\pm$ 2,045}}
        & \textcolor{red}{\textbf{64.55 $\pm$ 14.10}}
        & \$0.200 $\pm$ 0.031
        & $24\,\text{--}\,26$
        & \textcolor{red}{\textbf{10.08 $\pm$ 0.23}} \\

        DeepSeek-V4-Pro
        & 86,987 $\pm$ 13,725
        & 1525.33 $\pm$ 278.37
        & \textcolor{red}{\textbf{\$0.160 $\pm$ 0.024}}
        & $23\,\text{--}\,26$
        & 10.80 $\pm$ 0.26 \\
        \bottomrule
    \end{tabularx}
\end{table*}

{

\subsubsection{Ablation Study on Feature-Reduction Components}

We first conducted a component-level ablation study to evaluate the contributions of the three feature-reduction components in AutoDP-LLM: \emph{Semantic Guard}, \emph{Semantic Pruning}, and \emph{RF-MI Completion}. As shown in Table~\ref{tab:feature_reduction_ablation}, XGBoost and MLP were used as fixed representative classifiers for UNSW-NB15 and NSL-KDD, respectively, so that the observed differences could be attributed to changes in the feature-reduction configuration rather than to the choice of downstream classifier. Because Semantic Pruning operates on the semantic information established by Semantic Guard, removing Semantic Guard also disables Semantic Pruning, leaving RF-MI Completion as the only active component in that variant.

Removing RF-MI Completion caused the largest performance degradation on UNSW-NB15, reducing the weighted $F_1$-score from $78.41 \pm 0.22\%$ to $77.28 \pm 0.17\%$. A similar reduction was observed on NSL-KDD, where the score decreased from $76.38 \pm 0.99\%$ to $75.03 \pm 0.47\%$. These decreases of 1.13 and 1.35 percentage points, respectively, indicate that RF-MI Completion plays an important role in supplementing the semantically selected subset with statistically informative and non-redundant features.

When Semantic Pruning was removed while Semantic Guard and RF-MI Completion were retained, the weighted $F_1$-score decreased to $77.90 \pm 0.10\%$ on UNSW-NB15 and $76.26 \pm 1.14\%$ on NSL-KDD. The corresponding reductions of 0.51 and 0.12 percentage points suggest that Semantic Pruning provides an additional performance benefit, although its contribution is smaller than that of RF-MI Completion under the evaluated configurations.

Removing the entire semantic branch, including both Semantic Guard and Semantic Pruning, produced different effects across the two datasets. On UNSW-NB15, the RF-MI-only variant achieved $78.33 \pm 0.00\%$, only 0.08 percentage points below the full method. In contrast, its performance on NSL-KDD decreased substantially to $74.53 \pm 0.00\%$, representing a reduction of 1.85 percentage points. This result suggests that the benefit of semantic guidance is dataset-dependent and particularly pronounced on NSL-KDD. Nevertheless, the full configuration achieved the highest mean weighted $F_1$-score on both datasets, reaching $78.41 \pm 0.22\%$ on UNSW-NB15 and $76.38 \pm 0.99\%$ on NSL-KDD. Overall, the results support the complementary design of the semantic and statistical components: semantic processing preserves and refines domain-relevant information, whereas RF-MI Completion supplements the resulting subset using data-driven relevance and redundancy criteria.

\subsubsection{Ablation Study on LLM Backbones}

We further evaluated the influence of the underlying LLM backbone by comparing Claude-Sonnet-4.6, DeepSeek-V4-Pro, and GPT-5.4. Tables~\ref{tab:full_metrics_comparison_unsw} and~\ref{tab:full_metrics_comparison_nsl} report the downstream classification performance of the generated pipelines, while Table~\ref{tab:llm_backbone_comparison} summarizes their design-time overhead and runtime efficiency. For a consistent comparison, XGBoost and MLP were used as the fixed representative classifiers for UNSW-NB15 and NSL-KDD, respectively. Output tokens represent the number of tokens generated during pipeline construction, API latency measures the wall-clock time required to receive the complete generated pipeline, and cost denotes the corresponding API expenditure based on each provider's pricing. All three LLM engines successfully generated executable pipelines within the proposed multi-agent framework.

Under the fixed XGBoost configuration on UNSW-NB15, GPT-5.4 achieved the highest weighted precision and weighted $F_1$-score among the evaluated LLM engines, reaching $82.84 \pm 0.83\%$ and $78.41 \pm 0.22\%$, respectively. Its accuracy and weighted recall were both $77.20 \pm 0.23\%$, only 0.03 percentage points below the corresponding values obtained by DeepSeek-V4-Pro. By comparison, DeepSeek-V4-Pro achieved a weighted $F_1$-score of $78.34 \pm 0.11\%$, whereas Claude-Sonnet-4.6 obtained $77.44 \pm 0.18\%$. Although DeepSeek-V4-Pro with Random Forest produced the highest individual weighted $F_1$-score in the complete UNSW-NB15 comparison ($78.56 \pm 0.08\%$), GPT-5.4 provided the strongest overall result under the fixed representative-classifier setting.

On NSL-KDD, the advantage of GPT-5.4 was more pronounced under the fixed MLP configuration. It achieved the highest values across all four performance metrics, including accuracy and weighted recall of $78.80 \pm 0.92\%$, weighted precision of $82.63 \pm 0.83\%$, and a weighted $F_1$-score of $76.38 \pm 0.99\%$. In contrast, the MLP pipelines generated by Claude-Sonnet-4.6 and DeepSeek-V4-Pro obtained weighted $F_1$-scores of $74.82 \pm 0.40\%$ and $75.19 \pm 1.10\%$, respectively. GPT-5.4 therefore improved the mean weighted $F_1$-score by 1.56 percentage points over Claude-Sonnet-4.6 and by 1.19 percentage points over DeepSeek-V4-Pro under this configuration.

GPT-5.4 also demonstrated the highest pipeline-generation efficiency on both datasets. It required the fewest output tokens, with $10{,}418 \pm 2{,}139$ tokens on UNSW-NB15 and $10{,}800 \pm 2{,}045$ tokens on NSL-KDD. More notably, it achieved the lowest API latency, requiring only $60.23 \pm 8.44$~s and $64.55 \pm 14.10$~s, respectively. Claude-Sonnet-4.6 required $283.26 \pm 99.65$~s on UNSW-NB15 and $306.32 \pm 74.46$~s on NSL-KDD, approximately 4.7 times the latency of GPT-5.4 on both datasets. DeepSeek-V4-Pro exhibited substantially higher latencies of $858.33 \pm 354.50$~s and $1{,}525.33 \pm 278.37$~s, approximately 14.3 and 23.6 times those of GPT-5.4, respectively. Although DeepSeek-V4-Pro incurred the lowest monetary cost, GPT-5.4 remained less expensive than Claude-Sonnet-4.6 while providing substantially lower generation latency.

The pipelines generated by the three LLM engines retained 26--29 features on UNSW-NB15 and 23--27 features on NSL-KDD. Prior IDS feature-selection studies on UNSW-NB15 have reported compact feature subsets around this scale. Specifically, \cite{Tawfik2024} reports a subset containing 26 features for multiclass intrusion detection, while \cite{Almomani2020} reports selected subsets containing 25 and 30 features under different feature-selection procedures. These comparisons are intended only to contextualize the resulting dimensionality and do not indicate that a particular feature range is universally optimal.

When the same fixed selection thresholds were applied to NSL-KDD without further adjustment, the generated pipelines retained 23--27 features across the three LLM engines, including 24--26 features for GPT-5.4. These feature counts emerged from the semantic, relevance, and redundancy criteria rather than from a predefined target imposed by the framework. Their effectiveness is therefore assessed through the downstream classification results in Tables~\ref{tab:full_metrics_comparison_unsw} and~\ref{tab:full_metrics_comparison_nsl} and the component-level ablation results in Table~\ref{tab:feature_reduction_ablation}, rather than inferred from dimensionality alone.

GPT-5.4 also achieved the shortest training time with both representative classifiers: $13.55 \pm 0.45$~s with XGBoost on UNSW-NB15 and $10.08 \pm 0.23$~s with MLP on NSL-KDD. These results indicate that the choice of LLM backbone has its largest effect on the one-time design-time overhead, while the generated pipelines remain broadly comparable in dimensionality and runtime requirements. Overall, GPT-5.4 offered the most favorable balance of downstream predictive performance, generation efficiency, training efficiency, and monetary cost. Its highest weighted $F_1$-score under both fixed representative-classifier configurations, together with its substantially lower API latency, supports its selection as the representative LLM backbone in the subsequent experiments.

}

\subsubsection{Summary and Discussion}

{

The experimental results provide three main observations regarding the proposed AutoDP-LLM framework.

\begin{itemize}

\item \textbf{Domain-Guided Feature Reduction:}
AutoDP-LLM combines domain-guided semantic recommendations with TRAIN-derived Random-Forest and mutual-information evidence. The generated pipelines retain 26--28 features on UNSW-NB15 and 24--26 features on NSL-KDD while maintaining competitive multiclass classification performance. The component-level ablation results further show that RF--MI Completion contributes to performance on both datasets, whereas the observed benefit of semantic guidance is more pronounced on NSL-KDD.

\item \textbf{Comparison with Baseline Methods:}
AutoDP-LLM reports the highest weighted-$F_1$ value in the baseline comparison on both datasets. Its mean exceeds the strongest listed baseline point estimate by 0.24 percentage points on UNSW-NB15 and by 2.79 percentage points on NSL-KDD. However, the runtime comparison is method-dependent: AutoDP-LLM is faster than some baselines on UNSW-NB15 but does not provide a consistent runtime advantage across both datasets. The baseline methods produced identical classification metrics under the fixed evaluation protocol and are therefore reported as single-point estimates. Since no inferential comparison of paired test-set predictions was performed, the reported differences are interpreted descriptively.

\item \textbf{Evaluation Across Datasets and LLM Backbones:}
All three evaluated LLM backbones generated executable pipelines for both datasets. DeepSeek-V4-Pro with Random Forest obtained the highest individual mean weighted $F_1$-score on UNSW-NB15, while GPT-5.4 with MLP achieved the highest result on NSL-KDD. GPT-5.4 also required the fewest output tokens and the lowest mean API latency under the representative-classifier configurations, whereas DeepSeek-V4-Pro incurred the lowest monetary cost. These results demonstrate operational portability across the evaluated backbones while revealing different performance, latency, and cost trade-offs. The same core architecture was applied to UNSW-NB15 and NSL-KDD without dataset-specific structural modifications, providing initial evidence that the framework can accommodate different schemas and class distributions.

\end{itemize}

Taken together, the results show that AutoDP-LLM can automate IDS preprocessing and feature reduction while producing compact and executable pipelines. Its practical benefits depend on the dataset, downstream classifier, and selected LLM backbone.
}

\vspace{-0.1in}

\section{Conclusion and Future Work}

{

This paper presented and evaluated AutoDP-LLM, a multi-agent framework that uses LLMs to generate data-preprocessing and feature-reduction pipelines for intrusion-detection datasets. By organizing dataset analysis, transformation and bounded resampling, and feature reduction into specialized stages, the proposed framework automates several parts of IDS data preparation and reduces the manual effort required to design preprocessing workflows.

A key contribution of this work is the integration of domain-guided semantic recommendations with TRAIN-derived statistical evidence in an executable pipeline-generation framework. Experimental evaluations on the UNSW-NB15 and NSL-KDD benchmark datasets showed that the generated pipelines retained compact feature sets while achieving competitive multiclass intrusion-detection performance. Applying the same core framework to both datasets without changing its multi-agent architecture provides initial evidence of adaptability to different feature schemas and class distributions. In the representative GPT-5.4 experiments, the highest mean weighted $F_1$-scores were $78.41\%$ on UNSW-NB15 and $76.38\%$ on NSL-KDD. AutoDP-LLM also reported the highest weighted-$F_1$ value in the baseline comparison on both datasets, although its runtime advantage was not consistent across all methods.

The component-level ablation results indicated that RF--MI Completion contributed to predictive performance on both datasets, while the effect of semantic guidance was more pronounced on NSL-KDD. In addition, all three evaluated LLM backbones generated executable pipelines for both datasets. Their results revealed different trade-offs among predictive performance, generation latency, output-token use, training time, and monetary cost. These findings suggest that AutoDP-LLM provides a practical approach to automating IDS preprocessing, although its benefits depend on the dataset, downstream classifier, and selected LLM backbone.

Several research directions remain open. First, the applicability of the framework to more complex and less structured security data sources, such as raw packet captures and heterogeneous system logs, warrants further investigation. Second, future work will examine multicore, distributed, and HPC-enabled implementations to reduce pipeline-generation latency and support large-scale pipeline exploration. Third, smaller open-weight LLMs deployed locally should be investigated to reduce dependence on external APIs and support privacy-sensitive environments. Finally, extending the framework to continuous data streams may enable adaptive preprocessing under evolving network conditions and emerging attack patterns.
}

\vspace{-0.1in}

\FloatBarrier

\begin{appendices}
{

\begin{table*}[b]
\caption{Hyperparameter Configurations and Design Rationales for Multi-Agent Execution}
\label{tab:agent_hyperparams}
\centering
\renewcommand{\arraystretch}{1.4} 
\begin{tabularx}{\textwidth}{llcl}
\toprule
\textbf{Step / Agent} & \textbf{Parameter} & \textbf{Value} & \textbf{Design Rationale \& Objective} \\
\midrule
Step 1: Analysis \& Host Planning & Temperature & 0.0 & Promotes consistent planning and strict JSON schema compliance. \\
Step 2: Transformation \& Resampling Specialist & Temperature & 0.0 & Eliminates syntax hallucinations and guarantees deterministic code execution. \\
Step 3: Feature-Reduction Specialist & Temperature & 0.0 & Ensures consistent feature selection and deterministic reduction logic. \\
\bottomrule
\end{tabularx}
\end{table*}

\section{Execution Configurations, Hyperparameters, and System Prompt Templates}
\label{app:system_prompts}

{To support reproducibility, we report the requested LLM model identifiers and the generation parameters explicitly configured for each stage. The evaluated configurations corresponded to GPT-5.4, Claude-Sonnet-4.6, and DeepSeek-V4-Pro. Initial generation attempts used a temperature of 0.0 and a generation seed of 42. When an output failed contract validation, each correction attempt started from an isolated conversation, used a temperature of 0.1, and incremented the generation seed, beginning with 43 for the first correction attempt. The number of permitted correction attempts for each stage is described in Subsection~III-C. Generation parameters not explicitly supplied by the experimental client retained their provider defaults. These settings reduce avoidable output variation but do not guarantee identical responses from hosted LLM services.}

{

This appendix documents the complete execution configurations, hyperparameter settings, prompt templates, and input serialization formats of the AutoDP-LLM framework to facilitate reproducibility and experimental transparency. The hyperparameter configurations, including the specific temperatures and design rationales for each step within our workflow, are summarized in Table~\ref{tab:agent_hyperparams}. The three steps correspond directly to the three stacked blocks shown in Fig.~\ref{fig:figure2}: Step 1 is \emph{Analysis and Host Planning}, Step 2 is the \emph{Transformation and Resampling Specialist Agent}, and Step 3 is the \emph{Feature-Reduction Specialist Agent}. Each agent is executed as an independent LLM call, resulting in a multi-agent workflow rather than a single extended conversation.

\subsubsection{Shared Base Context (All Agents)} Every agent receives the following base context at the beginning of its system prompt, with the dataset name substituted at runtime and an optional prior preprocessing block appended when applicable, as illustrated in Fig.~\ref{fig:base_context}. This prior preprocessing block corresponds to the fixed, hand-written Data Ingestion and Profiling stage, which produces the \emph{Sanitized Dataset} that enters the workflow, as illustrated in Fig.~\ref{fig:figure2}.

\subsubsection{Analysis and Host Planning (Step 1)}
\label{subapp:analyst_system}

This step corresponds to the topmost block in Fig.~\ref{fig:figure2} (\emph{Analysis and Host Planning}) and is responsible for three sub-functions: \emph{Train-Only Profiling}, \emph{Resampling-Family Decision}, and \emph{Deterministic Plan Construction}, guided by the complete Step 1 system prompt in Fig.~\ref{fig:analyst_system}.

\subsubsection{Transformation and Resampling Specialist Agent (Step 2)}
\label{subapp:engineer_system}

This step corresponds to the middle block in Fig.~\ref{fig:figure2} (\emph{Transformation \& Resampling Specialist Agent}). It consumes the Validated Preprocessing Plan and generates the transformation and resampling function that implements the \emph{Guarded Drops \& Imputation}, \emph{Transformation, Encoding \& Scaling}, and \emph{Bounded Candidate Resampling} sub-functions based on the complete Step 2 system prompt detailed in Fig.~\ref{fig:engineer_system}.

\subsubsection{Feature-Reduction Specialist Agent (Step 3)}
\label{subapp:feature_system}

This step corresponds to the bottom block in Fig.~\ref{fig:figure2} (\emph{Feature-Reduction Specialist Agent}). It is responsible for three sub-functions: \emph{Semantic Guard \& Feature Lineage}, \emph{Evidence-Guided Semantic Pruning}, and \emph{Threshold-Based RF-MI Completion}. The output of this stage is the feature-reduction function, generated based on the complete Step 3 system prompt detailed in Fig.~\ref{fig:feature_system}.
}

\begin{figure*}[t]
\centering
\begin{adjustbox}{max width=\textwidth}
\begin{minipage}{\textwidth}

\begin{tcolorbox}[
    colback=gray!5!white,    
    colframe=gray!15!white,  
    arc=5pt,                 
    boxrule=0.5pt,           
    left=5pt, right=5pt, top=5pt, bottom=5pt 
]
\begin{lstlisting}[
    basicstyle=\scriptsize\ttfamily,
    breaklines=true,     % Tự động xuống dòng khi text quá dài
    columns=fullflexible
]
You are part of a multi-agent engineering team building a real-time, high-throughput intrusion detection system (IDS). Optimize preprocessing and feature reduction for strong detection quality and low inference latency. The primary selection metric is weighted F1 on validation data derived only from TRAIN. Report only weighted F1, accuracy, and per-class recall; accuracy and per-class recall are diagnostics.

Fit every learned statistic, transform, encoder, resampler, selector, and hyperparameter decision from TRAIN only. Held-out TEST labels, TEST class priors, and TEST feature distributions are unavailable for selection. TEST may only receive transformations already fitted or decided from TRAIN.
\end{lstlisting}
\end{tcolorbox}

\end{minipage}
\end{adjustbox}
\caption{Shared base context for all agents.}
\label{fig:base_context}
\end{figure*}

\begin{figure*}[b]
\centering
\begin{adjustbox}{max width=\textwidth}
\begin{minipage}{\textwidth}

\begin{tcolorbox}[
    colback=gray!5!white,    
    colframe=gray!15!white,  
    arc=5pt,                 
    boxrule=0.5pt,           
    left=5pt, right=5pt, top=5pt, bottom=5pt
]
\begin{lstlisting}[
    basicstyle=\scriptsize\ttfamily,
    breaklines=true,     % Tu dong xuong dong khi text qua dai
    columns=fullflexible
]
YOUR ROLE: Data Analyst Agent (Stage 1 of 3)

Produce one strict JSON decision; do not write code. Your only decision is exactly one TRAIN-only resampling family (or None). Do not select a winning candidate. Python owns cleaning, mild/moderate construction and the downstream controller on the same rare-safe TRAIN-derived split. That controller evaluates the entire none/mild/moderate grid using only standard LightGBM (`LGBMClassifier(random_state=42, verbosity=-1)`) and locks one shared winner for every final classifier.

RESAMPLING DECISION CONTRACT

- Return exactly one resampling_family string. Allowed values are None, RandomOverSampler, SMOTE, ADASYN, RandomUnderSampler, TomekLinks, EditedNearestNeighbours, SMOTETomek, and SMOTEENN.
- Never emit resampling_candidates, candidate IDs, intensities, sampler parameters, strategies, or budgets. Python constructs mandatory none and same-family mild/moderate candidates after validation.
- Preserve every class. Never merge, relabel, discard, or intentionally remove a rare class, including classes below one percent.
- Never match every class to the majority and never use an unrestricted or "auto" oversampling strategy.
- For oversampling/hybrid candidates, mild uses alpha=0.25 and a 1.05 row cap; moderate uses alpha=0.50 and a 1.10 row cap. A class is eligible only when its TRAIN count is below 0.5 times the median TRAIN class count. Its unscaled final target is ceil(n_c + alpha * (median - n_c)); scale additions proportionally when their sum exceeds the row cap.
- None uses no strategy. Pure under-sampling/cleaning uses alpha=null, max_expansion_ratio=1.0 and max_resampled_rows=TRAIN rows. The host supplies exact under_fraction, target_class_policy, method_params and strategy fields; Stage 1 does not emit them.
- Synthetic-neighbor methods require at least two original samples in every targeted class and an adaptive neighbor count below the smallest targeted support; otherwise prefer RandomOverSampler.
- Use TRAIN imbalance, absolute support, encoded categorical geometry, compute/memory cost, and compatibility with the downstream semantic-RF-MI selector. Large row count alone is not a reason to resample; mild imbalance with adequate support should normally return only "none".

Python derives conservative cleaning drops deterministically; Stage 1 must not emit cleaning decisions.

Python owns every deterministic budget, selection, encoding, scaling, transform, null-handling and audit field. Do not emit or redesign them. Use TRAIN evidence only. Return exactly one valid JSON object with no Markdown fence, comments, trailing commas, prose, or fields outside the user-provided decision schema.
\end{lstlisting}
\end{tcolorbox}

\end{minipage}
\end{adjustbox}
\caption{System prompt of Analysis Specialist Agent in Step 1.}
\label{fig:analyst_system}
\end{figure*}


\begin{figure*}[b]
\centering
\begin{adjustbox}{max width=\textwidth}
\begin{minipage}{\textwidth}

\begin{tcolorbox}[
    colback=gray!5!white,
    colframe=gray!15!white,
    arc=5pt,
    boxrule=0.5pt,
    left=5pt, right=5pt, top=5pt, bottom=5pt
]
\begin{lstlisting}[
    basicstyle=\scriptsize\ttfamily,
    breaklines=true,
    columns=fullflexible
]
YOUR ROLE: Code Engineer Agent (Stage 2 of 3)

Translate the approved Stage-1 plan into one production-grade Python implementation. The user prompt is the authoritative executable contract: follow its exact constants, API, operation order, categorical lineage, TRAIN-only fitting, bounded candidate-specific resampling, audits, and postconditions.

Do not redesign the plan, select a resampling winner, use validation/TEST labels or TEST-derived statistics, add dependencies or fallback encoders, or change candidate metadata. Use sklearn.preprocessing.TargetEncoder as required, preserve every TRAIN class, and return only one Python code block with no surrounding prose.

Keep the module concise: omit explanatory comments and duplicate self-audits; emit only the complete executable code and required contract audits so the response does not truncate.
\end{lstlisting}
\end{tcolorbox}

\end{minipage}
\end{adjustbox}
\caption{System prompt of Transformation and Resampling Specialist Agent in Step 2.}
\label{fig:engineer_system}
\end{figure*}


\begin{figure*}[t]
\centering
\begin{adjustbox}{max width=\textwidth}
\begin{minipage}{\textwidth}

\begin{tcolorbox}[
    colback=gray!5!white,
    colframe=gray!15!white,
    arc=5pt,
    boxrule=0.5pt,
    left=5pt, right=5pt, top=5pt, bottom=5pt
]
\begin{lstlisting}[
    basicstyle=\scriptsize\ttfamily,
    breaklines=true,
    columns=fullflexible
]
YOUR ROLE: Feature-Reduction Specialist Agent (Stage 3 of 3)

Implement the dataset-agnostic Semantic-RF-MI feature selector specified by the user prompt. Treat it as a behavioral contract for leakage-safe lineage, tiered protected semantic groups, TRAIN-only RF/MI evidence, adaptive K, minimal consumer audit metadata, and output correctness.

Do not use a benchmark-specific feature list, fuzzy/regex/keyword selection, validation or TEST labels/statistics, or ordinal target-code correlations. The host fits supervised feature-selection evidence once on the unresampled candidate `none`, then locks that feature list for all resampling candidates.
\end{lstlisting}
\end{tcolorbox}

\end{minipage}
\end{adjustbox}
\caption{System prompt of Feature Reduction Specialist Agent in Step 3.}
\label{fig:feature_system}
\end{figure*}


\section{JSON Schema for Inter-Agent Pre-processing Plan Transmission}
\label{app:json_schema}

{

The complete data-exchange contract between the \textit{Analysis and Host Planning step} and the subsequent downstream execution steps, corresponding to the \emph{Validated Preprocessing Plan} handoff illustrated in Fig.~\ref{fig:figure2}, is defined by the schema detailed in Fig.~\ref{fig:json_schema}. The Analysis agent first returns a strict single-field JSON decision specifying the resampling family. The Host Planning component validates this decision and deterministically synthesizes the complete Validated Preprocessing Plan, which constitutes the final output of Stage 1. Relevant plan fields and authoritative execution constants are then supplied to the downstream agents within fixed delimiter strings to preserve state integrity.
}

\begin{figure*}[b]
\centering
\begin{tcolorbox}[
    colback=gray!5!white,
    colframe=gray!20!white,
    arc=4pt,
    boxrule=0.5pt,
    left=6pt, right=6pt, top=4pt, bottom=4pt
]
\begin{minipage}[t]{0.485\textwidth}
\begin{lstlisting}[
    basicstyle=\scriptsize\ttfamily,
    breaklines=true,
    columns=fullflexible
]
{
  "rare_class_policy": "keep_all",
  "train_reference_rows": "<integer>",
  "train_reference_class_counts": {
    "class_label": "<integer>"
  },
  "resampling_selection": {
    "validation_source": "train_derived_validation",
    "splitter": "rare_safe_stratified_holdout",
    "validation_fraction": 0.20,
    "random_state": 42,
    "primary_metric": "weighted_f1",
    "weighted_f1_near_tolerance": 0.003,
    "tie_breakers": ["lower_intensity"],
    "refit_winner_on_full_train": false,
    "test_evaluations": 1
  },
  "resampling_candidates": [
    {
      "candidate_id": "none | mild | moderate",
      "imbalance_handling": "None | SMOTE | ...",
      "alpha": "<number_or_null>",
      "under_fraction": "<number_or_null>",
      "target_class_policy": "null | below_half_median | ...",
      "method_params": { "parameter": "<value>" },
      "max_expansion_ratio": "<float>",
      "max_resampled_rows": "<integer>",
      "sampling_strategy_dict": {
        "class_label": "<target_count>"
      },
      "justification": "..."
    }
  ],
  "resampling_budget_justification": "...",
  "downstream_compatibility_justification": "..."
  "imbalance_severity": "Mild | Moderate | Severe",
  "recommended_metrics": [
    "Weighted F1", "Accuracy", "Per-class Recall"
  ],
  "null_handling": {
    "feature": "median | mode | not_applicable_dropped"
  },
  \end{lstlisting}
\end{minipage}\hfill
\begin{minipage}[t]{0.485\textwidth}
\begin{lstlisting}[
    basicstyle=\scriptsize\ttfamily,
    breaklines=true,
    columns=fullflexible
]
  "drop": ["feature"],
  "drop_evidence": {
    "feature": {
      "reason_code": "constant | all_missing | ...",
      "evidence": "..."
    }
  },
  "categorical_encoding_policy": {
    "binary": "zero_one",
    "low_cardinality_max": 16,
    "low_cardinality_method": "onehot",
    "onehot_unknown_strategy": "explicit_unknown_bucket",
    "onehot_unknown_token": "__UNKNOWN__",
    "high_cardinality_method": "cross_fitted_multiclass_target",
    "target_encoder_cv": 5,
    "target_encoder_smooth": "auto",
    "target_encoder_shuffle": true,
    "target_encoder_random_state": 42,
    "fit_scope": "train_only"
  },
  "scaling_policy": {
    "method": "StandardScaler",
    "scale_continuous_numeric": true,
    "scale_target_encoded": true,
    "scale_binary": false,
    "scale_onehot": false,
    "fit_before_resampling": true,
    "fit_scope": "train_only"
  },
  "log_transform_policy": {
    "method": "train_only_right_skew_rule",
    "minimum_unique_values": 20,
    "minimum_skewness": 2.0,
    "upper_tail_iqr_multiplier": 3.0
  },
  "transform": {},
  "encode": {
    "feature": {
      "train_cardinality": "<integer>",
      "method": "zero_one | onehot | ... "
    }
  },
  "normalization": "StandardScaler"
}
\end{lstlisting}
\end{minipage}
\end{tcolorbox}
\vspace{-2mm}
\caption{Schema of the synthesized Validated Preprocessing Plan.}
\label{fig:json_schema}
\end{figure*}


}
\end{appendices}

\FloatBarrier
\clearpage


\section*{Declarations}
\textbf{Competing interests} The authors declare no competing interests.

\textbf{Ethics approval} Not Applicable.

\textbf{Consent to participate} Not Applicable.

\textbf{Consent to publish} Not Applicable.

\textbf{Funding} The work of Thien Van Luong was supported by National Economics University under grant number NEU1-2025.02. The work of Thien Van Luong was also supported by Vietnam National Foundation for Science and Technology Development (NAFOSTED) under grant number 102.05-2025.57.

\bibliographystyle{IEEEtran}
\phantomsection\addcontentsline{toc}{section}{\refname}\bibliography{REFS}

%

\end{document}